\documentclass[fleqn,usenatbib]{mnras}

\usepackage{newtxtext,newtxmath}

\usepackage[T1]{fontenc}

\DeclareRobustCommand{\VAN}[3]{#2}
\let\VANthebibliography\thebibliography
\def\thebibliography{\DeclareRobustCommand{\VAN}[3]{##3}\VANthebibliography}

\usepackage{graphicx}	% Including figure files
\usepackage{amsmath}	% Advanced maths commands
\usepackage{subcaption}
\usepackage{cuted}
\usepackage{xcolor}
\usepackage{soul}
\usepackage{cuted}
\usepackage{afterpage}

\title[Using cross-correlations to probe BBH origins]{Connecting The Hierarchically Merging Binary Black Hole Population To Their Host Galaxies}
\author[Moncrieff, Panther]{
Jordan W.N. Moncrieff,$^{1,2}$\thanks{E-mail: jordan.moncrieff@research.uwa.edu.au}
Fiona H. Panther$^{1,2}$
\\
$^{1}$Department of Physics, University of Western Australia, Crawley WA 6009, Australia\\
$^{2}$OzGrav: The ARC Centre of Excellence for Gravitational-wave Discovery , Crawley WA 6009, Australia
}

\date{Accepted XXX. Received YYY; in original form ZZZ}

\pubyear{2025}

\begin{document}
\label{firstpage}
\pagerange{\pageref{firstpage}--\pageref{lastpage}}
\maketitle

\begin{abstract}
The detection of gravitational waves from merging black holes with masses $\sim\,80-150\,\mathrm{M_\odot}$ suggests that some proportion of black hole binary systems form hierarchically in dense astrophysical environments, as most stellar evolution models cannot explain the origin of these massive black holes through isolated binary evolution. A significant fraction of such mergers could occur in Active Galactic Nuclei disks (AGN), however connecting individual black hole mergers to host galaxies is a challenging endeavor due to large localization uncertainties. We assess the feasibility of determining the fraction of hierarchically merging black hole binaries by computing the angular cross-correlation between gravitational wave localization posteriors and galaxy catalog skymaps. We forecast when the clustering of gravitational wave sky localizations can be measured accurately enough to distinguish the AGN origin scenario from hierarchical mergers in galaxies that do not host AGN. We find that if the observed merging population is dominated by binaries formed dynamically in AGN, then this could be determined with $\mathcal{O}(5000)$ mergers detected at the sensitivity that is projected for the upcoming A\# gravitational wave detectors.
\end{abstract}


\section{Introduction}\label{sec:Intro}

Binary black holes (BBH) can form though isolated binary evolution, or hierarchical mergers in dense astrophysical environments \citep{escriva2024black}. Isolated BBH systems originate from stellar collapse in binary star systems \citep{mapelli2020binary}. However, the formation of black holes within the mass range $\sim 50-130 M_\odot$ \citep{chatzopoulos2012effects, belczynski2016effect} is restricted by the pair-instability process, which results in a remnant being left by the star's terminal supernova explosion. Consequently, the presence of the `upper mass gap' in the mass distribution of a black hole population serves as a key observational signature of isolated binary stellar evolution, although its precise lower bound remains debated \citep{ziegler2021filling, woosley2021pair, farag2022resolving}. Conversely, hierarchical mergers, where black holes form from repeated mergers in dense astrophysical environments, can populate this mass gap with second and higher-generation black holes \citep{o2006binary}. Analysis of the population of BBH detected by the LIGO-Virgo-KAGRA (LVK) collaboration in the Third Gravitational Wave Transient Catalog \citep[GWTC-3,][]{abbott2023population} shows no evidence of a suppressed BBH merger rate at mass $>60 M_{\odot}$, implying that the upper mass gap is occupied. In particular, GW190521 was the result of the merger of black holes with component masses $85^{+21}_{-14} M_{\odot}$ and $66^{+17}_{-18} M_{\odot}$ \citep{abbott2020properties,abbott2020gw190521}, both of which lie inside (most estimates of) the mass gap. This suggests that the black holes in GW190521 could have been formed through hierarchical mergers. 

Several astrophysical environments can facilitate such mergers, including AGN disks \citep{atallah2023growing, rose2022formation,tagawa2020formation,ford2022binary}, and dense star clusters \citep{rozner2022binary,mapelli2021hierarchical,rodriguez2019black}. Irrespective of the environment in which they form, hierarchically merging BBHs are expected to have different mass and spin distributions than the isolated binary channel \citep{gerosa2021hierarchical}, in principle allowing the two populations to be separated by analysis of the GW data \citep{antonini2025star}. Additionally, hierarchically merging binaries in AGN disks are expected to have different distributions in spins, mass ratios, eccentricities, etc, compared to hierarchical mergers in dense star clusters \citep{ford2022binary}, such as globular clusters and nuclear star clusters \cite{escriva2024black}.  However, determining whether hierarchically-formed binaries come primarily from AGN disks or dense star clusters is complicated by uncertainties in the astrophysics of dense environments \citep{li2024origin}, with the rates of mergers from each channel depending on unknown parameters for the efficiency of migration and mass segregation \citep{li2023comparing}.

Thus, while the population of hierarchically merging events may be readily distinguished from the field binary population in the coming years through the distributions of their intrinsic parameters, it is challenging to understand which formation channel -- if any --  dominates the rate of hierarchical mergers: AGN-driven mergers or dynamically driven mergers in dense star clusters.

Identifying how mergers in AGN contribute to the rate of hierarchically merging BBH is particularly interesting. Due to the dynamics of mergers in a gas-rich environment, it has been proposed that AGN can mediate a higher rate of BBH mergers \citep{ford2022binary} relative to dense star clusters. Moreover, the AGN channel has potential to explain very high mass events, like GW190521 \citep{abbott2020gw190521} and GW231123 \citep{ligo2025gw231123}, as the deep gravitational potential well of an AGN effectively retains merger products, allowing for multiple generations of mergers to occur \citep{tagawa2021mass}. However, these rates are highly uncertain \citep{tagawa2020formation}.

Associating individual BBH with host galaxies typically requires precise localization, which would allow us to determine whether hierarchical mergers do occur with a greater frequency in AGN disks than dense star clusters. However, even planned next-generation detectors such as Cosmic Explorer \citep{reitze2019cosmic} and LISA \citep{punturo2010einstein} are expected to enable localization of BBH mergers to regions of $\mathcal{O}(1) \rm{deg}^2$  \citep[$90\%$ credible interval][]{gupta2023characterizing}. Regions of this size can contain tens to hundreds of possible host galaxies, with the exception of only the highest SNR events, which are expected to be few in number \citep{mo2024identifying}.

An alternative method to precisely localise these mergers is via their putative electromagnetic counterparts \citep{kimura2021outflow,tagawa2023observable} or high-energy neutrino emissions \citep{zhou2023high,ma2024high}. Such a detection would enable multi-messenger astronomy with BBHs \citep{morton2023gw190521}. While electromagnetic counterparts are only expected from mergers in AGN, and hence could be used as a smoking gun signature of AGN origin to pinpoint host galaxies, statistical associations between AGN flares and BBH mergers, such as those proposed for GW190521 \citep{graham2020candidate, graham2023light}, remain inconclusive. However, because scenarios that result in the production of EM counterparts arising from the interaction between the merging black holes and the AGN disk are somewhat fine-tuned \citep{mckernan2019}, we may expect such coincidences to be the exception, not the rule.

We consider alternative methods to correlate hierarchically merging black holes with host galaxy populations. Different types of galaxies cluster in the sky differently. AGN are more strongly clustered in the universe due to their environmental preference for massive cluster-core galactic hosts \citep{seymour2007massive} and thus have a distinguishable clustering bias compared to galaxies that do not host AGN \citep{hale2018clustering}. This remains true even when accounting for galaxy mass and redshift \citep{banerjee2023clustering}. Thus, if BBHs merge primarily within AGN disks, their clustering properties should be comparable to that of AGN. By measuring this cross-correlation, we can statistically infer the contribution of AGN-hosted mergers to the overall BBH population. Although dense star clusters are also expected to exist inside of AGN, there is no known preference of star clusters to occur in AGN and therefore this small contribution would not change the clustering behavior noticeably.

In this work, we show how measuring the angular cross-correlations between the localization posteriors of a population of hierarchical BBH mergers and galaxy catalogs could allow us to deduce whether the AGN-driven or dense cluster formation scenario dominates the hierarchical merger rate. 

\section{cross-correlation of gravitational wave and galaxy skymaps}

\subsection{Angular clustering}
We wish to measure the degree to which galaxies in each of our catalogs cluster together on the plane of the sky -- that is, the over- or under-density of galaxies in a given region relative to the mean density of galaxies $\bar{N}$. For a given galaxy catalog containing $N(\theta,\phi)$ galaxies within a small region of the sky centered at right ascension $\theta$ and declination $\phi$, we can define the density contrast at that point as

\begin{equation}\label{eq:delta_def}
    \delta(\theta,\phi)=\frac{N(\theta,\phi)-\bar{N}}{\bar{N}}.
\end{equation}

\noindent
In practice $(\theta, \phi)$ are replaced by discrete pixels. 
Then, decomposing the density contrast $\delta(\theta,\phi)$ into a spherical harmonic basis,
\begin{equation}
    \delta(\theta,\phi)=\sum_{{\ell},m} a_{{\ell}m} Y_{{\ell}m}(\theta,\phi),
\end{equation}
\noindent
where the summation is performed over integers $\ell=0,1,...,\ell_{\rm{max}}$, up to some cut-off $\ell_{\rm{max}}$, and $-\ell \le m \le \ell$. The $a_{{\ell}m}$ coefficients can be found by integrating over the sphere $\Omega \equiv(\theta,\phi)$
\begin{equation}\label{eq:a_lm def}
    a_{\ell m} = \int  \delta(\Omega) \left[ Y_{\ell m}(\Omega) \right]^* d\Omega .
\end{equation}
\noindent
The angular cross-correlation (ACC) coefficients $C_{\ell}$
 are real numbers that represent the variance in the absolute value of the $a_{\ell m}$, i.e.

\begin{equation}\label{eq:Cl_defn}
    C_{\ell} := \frac{1}{2 \ell +1} \sum_m a_{\ell m} a_{\ell m}^*
\end{equation}

These coefficients are directly related to the clustering of matter on the sky \citep{magneville2007notes}. 

On angular scales large enough that non-linear effects are negligible \citep{desjacques2018large}, clustering can be well described by a single linear bias factor, with 
\begin{align}
C_{\ell}^{\rm{N}}=b_{\rm{N}} C_{\ell}^{\rm{matter}}, \quad C_{\ell}^{\rm{A}}=b_{\rm{A}}  C_{\ell}^{\rm{matter}}
\end{align}
where $C_\ell^{\rm{matter}}$ is the ACC of matter density in the universe, while A and N refer to AGN and non-AGN galaxy skymaps respectively. Then, the relative bias 
\begin{equation}
b \equiv b_{A} / b_{N}
\end{equation}
allows for a measure of the relative clustering of AGN and non-AGN without reference to the general matter distribution, with $C_\ell^{A}=b  C_\ell^{N}$.

In this work we wish to measure the angular cross-correlation a catalog $X$ with another catalog $Y$. This is given by:

\begin{equation}\label{eq:Cl_cross_def}
    C_{\ell}^{X,Y}=\frac{1}{2\ell + 1} \sum_{m} a_{\ell m}^X \cdot (a_{\ell m}^Y)^*,
\end{equation}

\noindent
where $a_{\ell m}^X$ and $a_{\ell m}^Y$ are the spherical harmonic expansion coefficients for the density contrasts between catalogs $X$ and $Y$ respectively. Measuring the cross-correlation in this way has the advantage of lowering shot noise, allowing a more accurate test of the clustering of a catalog containing relatively sparse objects, such as a gravitational wave catalog.

\subsection{Galaxy catalogs}\label{sec:catalogs}
\begin{figure}
    \centering
    \includegraphics[width=1.0\linewidth]{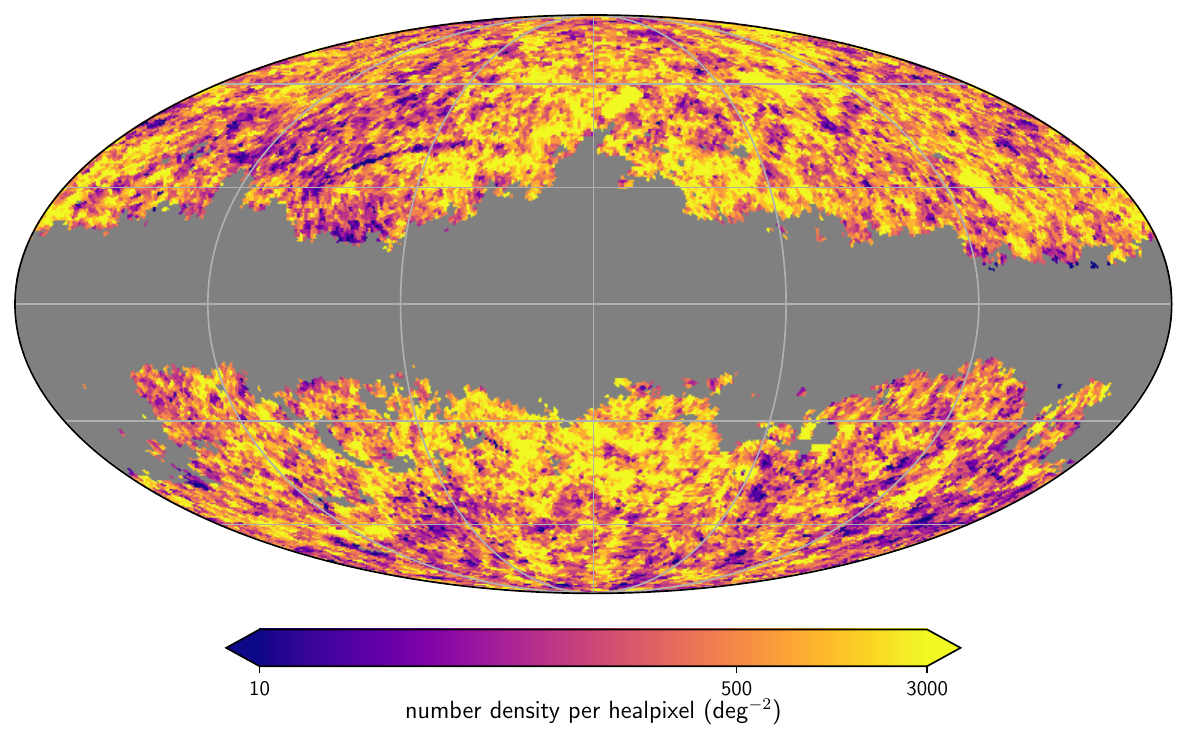}
    \caption{Masked galaxy pixel map from the GLADE+ catalog, where non star forming galaxies (including AGN) are removed.}
    \label{fig:glade+_catalog_sfg}
\end{figure}

\begin{figure}
    \centering
    \includegraphics[width=1.0\linewidth]{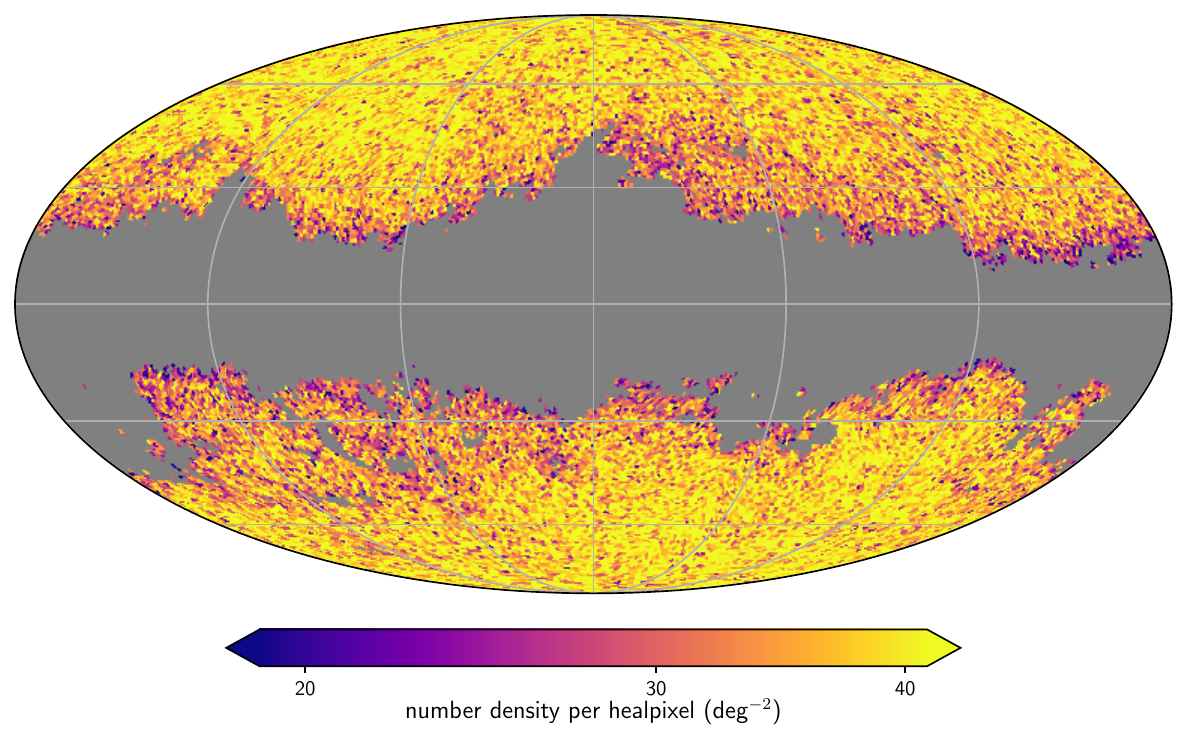}
    \caption{QUAIA quasar catalog, with the same mask used for GLADE+.}
    \label{fig:WISE_AGN_density_contrast}
\end{figure}

To compare how our simulated gravitational wave population clusters compared to galaxies, we use two galaxy catalogs. The scenario we present is idealized, in the sense that we assume that these catalogs are complete and a true representation of the distribution of galaxies that will be measured out to a comparable redshift to our gravitational wave population. Due to current limitations in galaxy surveys, which are typically designed to optimize completeness for a particular galaxy type, we choose to draw our AGN and non-AGN galaxies from two separate catalogs:
\begin{itemize}
\item We draw our non-AGN sample from the GLADE+ all-sky galaxy catalog \citep{dalya2022glade+}, which combines several different galaxy catalogs into one large catalog, to be used primarily for gravitational wave follow up observations. See Figure   \ref{fig:glade+_catalog_sfg}, showing the galaxy density field of the GLADE+ catalog using the Healpix scheme \citep{2005ApJ...622..759G} as implemented in the Python package Healpy \citep{Zonca2019}. 
\item We draw our AGN sample from the  QUAIA GAIA-unWISE all-sky photometric quasar catalog \citep{storey2024quaia}, which contains 1,295,502 million quasars\footnote{The catalog can be downloaded from \url{https://zenodo.org/records/10403370}.}, shown in Figure \ref{fig:WISE_AGN_density_contrast}.
\end{itemize}

These catalogs are chosen due to their high levels of completeness in the local universe as well as their all-sky coverage. We do note, however, that the specific choice of galaxy catalog used in this work does not significantly change our results. 

To ensure our non-AGN galaxy catalog has a high purity, we remove all identified quasars from the GLADE+ catalog. We limit our AGN sample to the local universe, removing all quasars with redshifts $z>1.5$ from our AGN sample. 

Many galaxy catalogs suffer from significant incompleteness close to the galactic plane due to dust extinction. For this reason, we mask regions on the sky map where the galaxy completeness is less than $0.5$ (e.g.  \cite{alonso2023constraining}). We find that although this mask excludes around 40 per cent of the sky, it does not significantly influence our results. We do not expect the clustering of AGN to evolve much in the local universe \citep{donoso2014angular}, and so we use a single redshift bin. Exploration of the use of 3D clustering is deffered to future work.

\begin{figure}
    \includegraphics[width=1.0\linewidth]{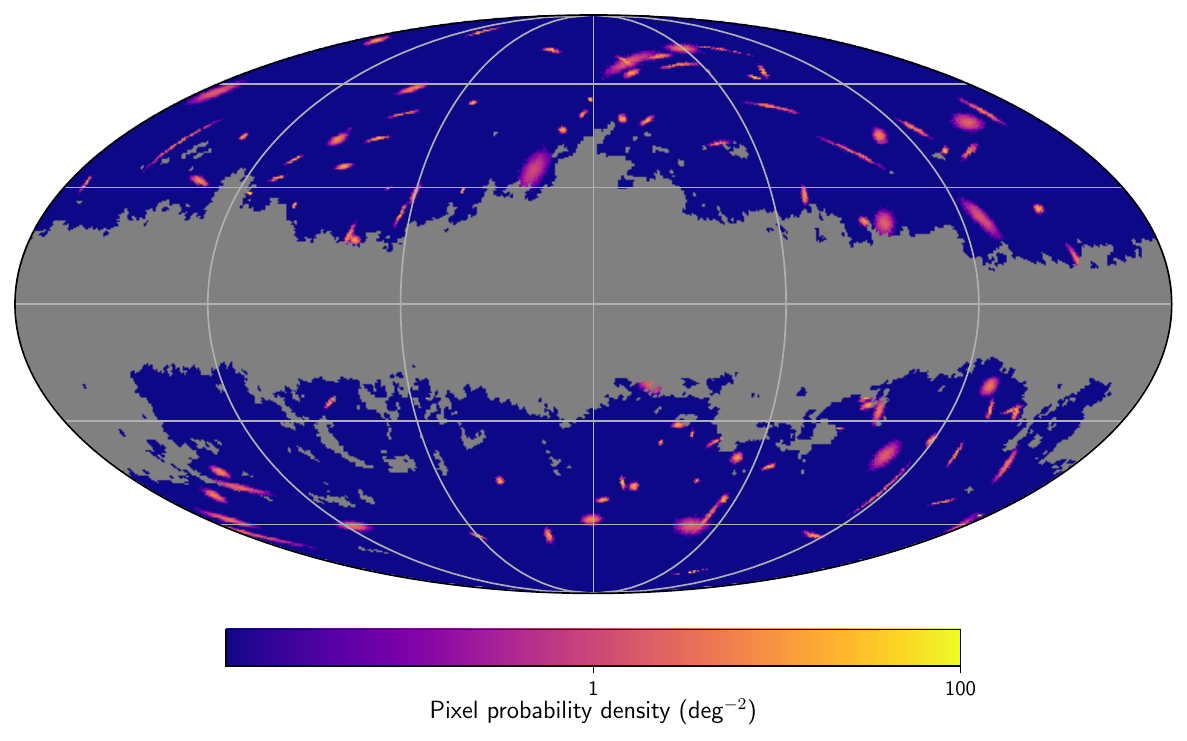}
    \caption{An example of $N=100$ GW sky localization posteriors for events with true locations sampled from the QUAIA quasar catalog. The sky localizations are computed using \texttt{gwbench}.}
    \label{fig:N=100_BBH_posteriors}
\end{figure}

\subsection{Binary black hole merger population}\label{sec:GW_skymap}

In this work we simulate the detection of BBHs with future detector networks to generate skymaps that will be cross-correlated with our galaxy catalogs. We can impose constraints on the localization uncertainty of BBH mergers by employing the Fisher Information Formalism (FIF) for signals with sufficiently high signal-to-noise ratio \citep{vallisneriFisher2008, RodriguezFisher2013}. By inverting the Fisher Information Matrix (FIM, $\Gamma$), we can obtain an estimate of the uncertainty in the right ascension and declination of our signals for arbitrary detector networks. We use the publicly available software package \texttt{gwbench} \citep{borhanian2021gwbench} to estimate the $1\sigma$ uncertainties in the sky localization for each of our BBH signal injections using the FIF. 

We simulate BBH signals using the TaylorF2 waveform approximant. The true sky locations are drawn from galaxies in the appropriate galaxy catalog. We assume a population of BBHs with independent masses that are uniformly distributed between $10-100$ solar masses, with aligned spins that are uniformly distributed $\chi_z \sim \rm{Unif}(0,1)$. While a truly hierarchically merging population is expected to have a different distribution of parameters, these do not impact the waveforms significantly enough to affect the recovered SNR, which dominates the determination of the sky localization area. We apply a  standard flat $\Lambda\mathrm{CDM}$ cosmology with redshifts $z \sim \rm{Unif}(0,1)$. We make this latter choice because we ignore any three-dimensional correlations in this work, instead focusing on clustering on the plane of the sky. Furthermore, in the case of our AGN catalog, the majority of the galaxy redshifts are somewhat uncertain \citep{storey2024quaia}. Incorporating the full 3D localizations for the GW events is deferred to future works. We use this BBH population to represent a future BBH population that is known to derive from hierarchical mergers. We take three estimates for the expected number of high-quality detections at A\# design sensitivity from \cite{broekgaarden2024visualizing} and simulate three populations containing, respectively, $N=[1000, 5000, 10000]$ BBH mergers.

These signals are injected into coloured Gaussian noise. The power spectral density used to colour the noise is defined by the detector network configuration. We simulate the recovery of signals at A\# design sensitivity \citep{lsc2022report}, with a detector network consisting of LIGO Hanford, LIGO Livingston, and VIRGO detectors. We use the A\# PSD supplied by the \texttt{gwbench} package \citep{borhanian2021gwbench}. From our $N$ simulated events, we remove any that are detected with a combined $\mathrm{SNR}>30$. We then select only events that are detected with $\mathrm{SNR}>12$ in all detectors. This removes $\sim 1-5\,\mathrm{per\,cent}$ of our simulated signals in each catalog of GW events, however it ensures the validity of the assumptions made in the FIF. 

An example of the population of sky localizations, obtained from \texttt{gwbench} for a BBH population of $N=100$ events with injected true localizations sampled from the quasar catalog, is shown Figure \ref{fig:N=100_BBH_posteriors}.

\subsection{Determining galactic hosts via a point estimate of the angular cross-correlations}\label{sec:ACC_point_estimate}

Given a gravitational wave skymap $GW$, and we want to determine the clustering of the GWs on the sky. In particular, we can compare the clustering of GWs with that of AGN versus non-AGN galaxies. In doing so, we can test the two hypothesis that the GWs are sourced from BBH mergers in AGN versus BBHs in non AGN galaxies. We can do this by cross correlating the GW skymaps separately with both an AGN and a non-AGN galaxy skymaps, and compare the resulting best-fitting bias parameters between the cross-correlations and the auto correlations. For example, if the GW catalog was sourced by locations in the AGN catalog, then the expected value of the cross-correlation $C_\ell^{\rm{GW},\rm{A}}$ is $C_\ell^{\rm{A},\rm{A}}$, where GW refers to the gravitational wave catalog and A refers to an AGN catalog. Additionally, the cross-correlation of this catalog with a non-AGN catalog N, $C_\ell^{\rm{GW},\rm{N}}$, would have an expected value given by $C_\ell^{\rm{A},\rm{N}}$. In contrast, if the GW skymap was generated by BBH mergers in non-AGN, then the expected value $C_\ell^{\rm{GW},\rm{N}}$ when cross-correlated with the non-AGN sky map is $C_\ell^{\rm{N},\rm{N}}$, and the expected value $C_\ell^{\rm{GW},\rm{A}}$ when cross-correlated with an AGN catalog is $C_\ell^{\rm{A},\rm{N}}$.

If we assume a particular formation hypothesis, say the AGN origin scenario, then we can estimate the error associated with our point estimates of $C_\ell^{\rm{GW},\rm{A}}$  $\left( C_\ell^{\rm{GW},\rm{N}} \right)$. The covariance matrix for the cross-correlation of a galaxy skymap CAT with a GW skymap with Gaussian localizations is given by \citep{mukherjee2022cross}:

\noindent
\begin{minipage}[t]{0.48\textwidth}
\rule[0pt]{\linewidth}{0.4pt}
\end{minipage}
\begin{strip}
\centering
\begin{equation}\label{eq:Covariance_matrix}
 \Sigma_{\ell \ell'}^{\rm{CAT},\rm{GW}}=\delta_{\ell \ell'}\bigg [\bigg(C_\ell^{\rm{CAT},\rm{CAT}}+\frac{4\pi f_{\rm{sky}}}{N_{\rm{CAT}}}\bigg)\bigg(C_\ell^{\rm{GW},\rm{GW}}+\frac{4\pi f_{\rm{sky}}}{N_{\rm{GW}}}\bigg)+\bigg(C_\ell^{\rm{GW},\rm{CAT}}\bigg)^2\bigg ] \bigg[(2 \ell+1)\cdot \Delta \ell \cdot f_{\rm{sky}}\bigg]^{-1},
\end{equation}
\end{strip}

\noindent
\begin{minipage}[t]{0.48\textwidth}
\rule[0pt]{\linewidth}{0.4pt}
\end{minipage}

\noindent
where $N_{\rm{CAT}}$ and $N_{\rm{GW}}$ are the number of objects in the galaxy and GW catalogs respectively, $f_{\rm{sky}}$ is the fraction of the sky contained in the skymaps (after masking), and $\Delta \ell$ is the number of the $C_\ell$ coefficients the ACC is averaged over. 

The expected distribution of $C_\ell^{\rm{GW},\rm{A}}$ ($C_\ell^{\rm{GW},\rm{N}}$), assuming no biases inherent in the GW detection, will be $\mathcal{N}(C_\ell^{\rm{A},\rm{A}},\Sigma_{\ell,\ell}^{\rm{GW},\rm{A}})$ $\left( \mathcal{N}(C_\ell^{\rm{N},\rm{A}},\Sigma_{\ell,\ell}^{\rm{GW},\rm{N}}) \right)$. On the other hand, if the GWs are sourced from the non-AGN catalog N, then the expected distributions are $C_\ell^{\rm{GW},\rm{A}} \sim \mathcal{N}(C_\ell^{\rm{N},\rm{A}},\Sigma_{\ell,\ell}^{\rm{N},\rm{GW}})$ and $C_\ell^{\rm{GW},\rm{N}} \sim \mathcal{N}(C_\ell^{\rm{N},\rm{N}},\Sigma_{\ell,\ell}^{\rm{GW},\rm{A}})$.

\begin{figure*}
    \centering
    \includegraphics[width=1.0\textwidth]{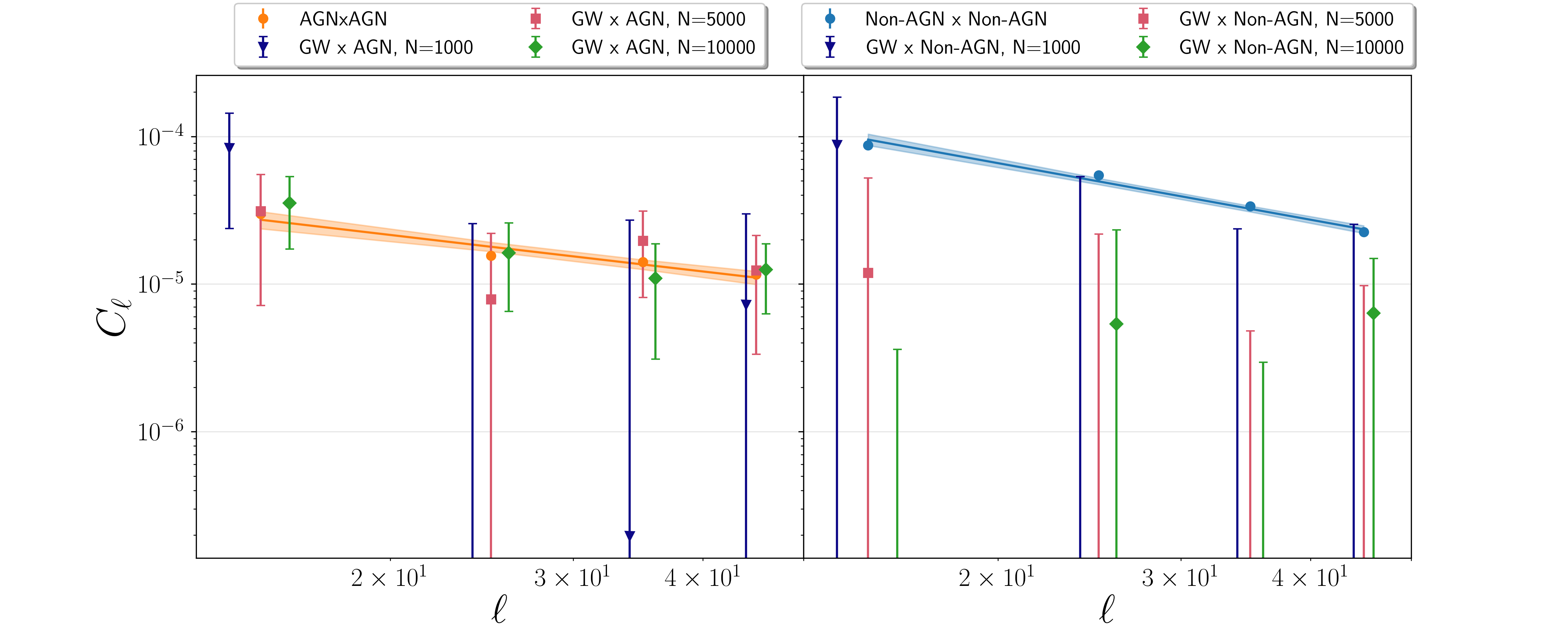}
    \caption{On the left hand side, we show that when the true BBH locations are sampled from the AGN catalog there is good agreement between the ACC computed between the GW catalog and AGN catalog (GW x AGN), and the autocorrelation of the AGN catalog (AGNxAGN, shown as orange circles, along with its power law best fit and the 90 per cent confidence interval region shaded). On the right hand side, we show that there is no support for the ACC between the GW catalog (with true locations drawn from an AGN catalog) and the non-AGN catalog being consistent with the autocorrelation of the non-AGN galaxy catalog. The $C_\ell$ coefficients plotted are band-averaged over the interval $\left[C_\ell-\Delta \ell/2,C_\ell+\Delta \ell/2 \right]$ with $\Delta \ell=10$; for visual clarity we offset the $N=1000$ $C_\ell$ coefficients on the plot. Error bars on the GW cross-correlations and the galaxy auto-correlations are given by Equation \ref{eq:Covariance_matrix}.}
    \label{fig:Cl_GWAGN_point_estimate}
\end{figure*}

\begin{figure*}
    \centering
    \includegraphics[width=1.0\textwidth]{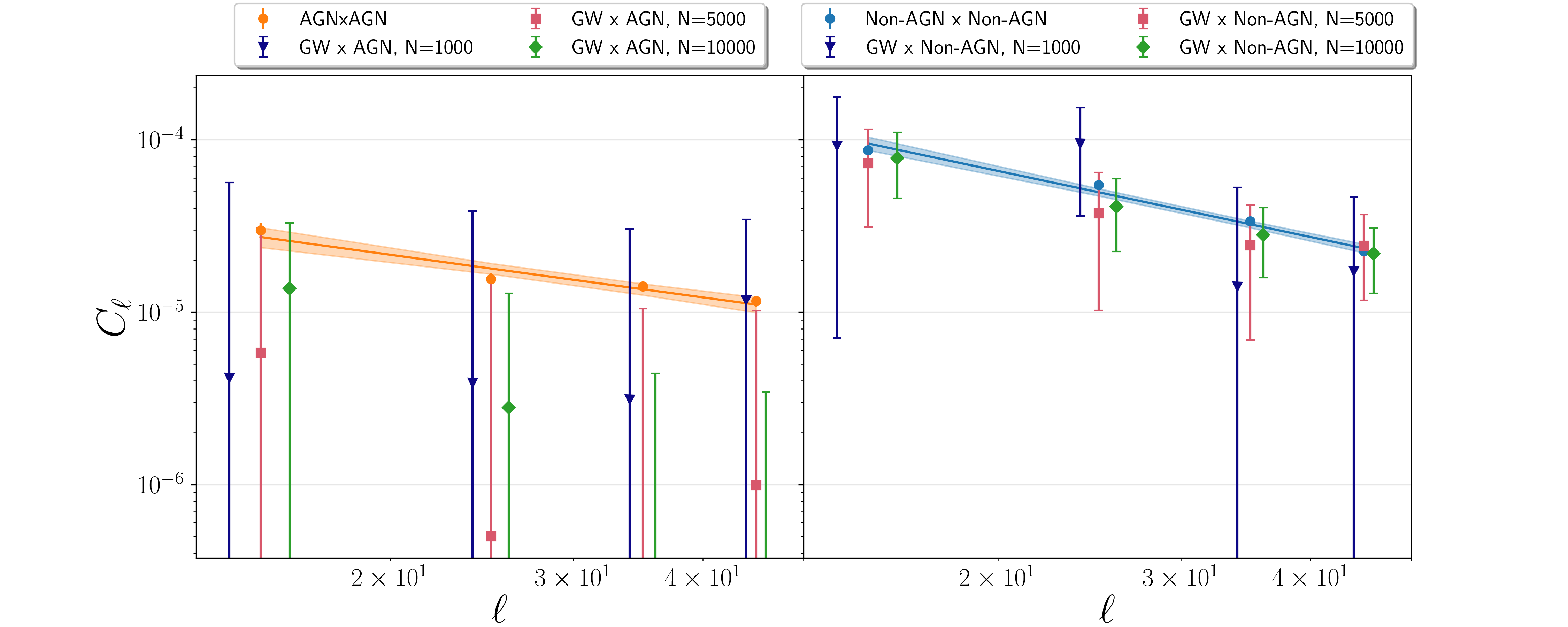}
    \caption{Same as Fig. \ref{fig:Cl_GWAGN_point_estimate}, but the true BBH locations sampled from locations in the non-AGN catalog. In this case there is greater support for the ACC between the GW catalog and non-AGN being consistent with the non-AGN catalog within statistical uncertainties, as expected.}
    \label{fig:Cl_GWSFG_point_estimate}

\end{figure*}

To demonstrate the method, we show six examples of inferring the galactic progenitor of GW catalogs via the angular cross-correlation. Figures \ref{fig:Cl_GWAGN_point_estimate} and \ref{fig:Cl_GWSFG_point_estimate} show the result of computing the ACC between the GW catalog and galaxy catalogs (generated from the AGN and non-AGN catalogs described in Section \ref{sec:catalogs}). Six different GW skymaps are used, the three in Fig. \ref{fig:Cl_GWAGN_point_estimate} are all generated with true locations drawn from an AGN catalog, 
while the three in Fig. \ref{fig:Cl_GWSFG_point_estimate} are drawn from the non-AGN catalog. For each case, we can infer the origin of the GWs in each catalog by computing the ACC of the GW skymaps with each of the AGN and non-AGN galaxy skymaps. For each figure we show the three examples with different numbers of BBHs, $N_{BBH}=1000$, $N_{BBH}=5000$, and $N_{BBH}=10,000$, showing how the error in the ACC decreases as more detections occur.

In Figure \ref{fig:Cl_GWAGN_point_estimate}, the point estimate of $C_\ell^{GW,X}$ - with $2 \sigma$ error-bars computed from Equation \ref{eq:Covariance_matrix} - are shown for each GW catalog. The $2\sigma$ error-bars are computed via $\Delta C_{\ell}^{AGN}=2\sqrt{\Sigma_{\ell \ell'}^{A,A}}$, where $\Sigma_{\ell \ell'}^{A,A}$ is found by setting $C_{\ell}^{CAT,CAT}=C_\ell^{GW,GW}=C_\ell^{GW,CAT}=C_\ell^{A,A}$ in Equation \ref{eq:Covariance_matrix}. For $N=1000$ events there is substantial scatter in the derived $C_\ell^{GW,A}$, with no clear agreement between these values and the AGN catalog autocorrelation (left hand side, orange) or the non-AGN catalog autocorrelation (right hand side, blue). As the number of events in the catalog increases, $C_\ell^{GW,A}$ converges toward $C_\ell^{A,A}$ (left hand side). We see no such trend when comparing $C_\ell^{GW,N}$ and $C_\ell^{A,A}$ (right hand side). We draw the conclusion that the AGN origin hypothesis is thus preferred over a non-AGN origin. Similarly, if our true GW locations are drawn from the non-AGN catalog, the same phenomenon is observed with good agreement between the autocorrelation of the non-AGN catalog and the ACC between the GW catalog and non-AGN catalog with 5000 events (Fig. \ref{fig:Cl_GWSFG_point_estimate}). The larger number of objects in the non-AGN catalog yields somewhat smaller errorbars.

\subsection{Cross correlating mock GW skymaps with galaxy catalogs}\label{sec:cross_cor_sim}

Having shown the practical procedure in the case of a single simulated GW catalogs, here we show that we can expect that our point estimates of the ACC is an unbiased estimator. We compute point estimates with many different GW catalog realizations, showing that the average values obtained by the point estimate indeed converge to the expected values.

\begin{figure}
    \includegraphics[width=\linewidth]{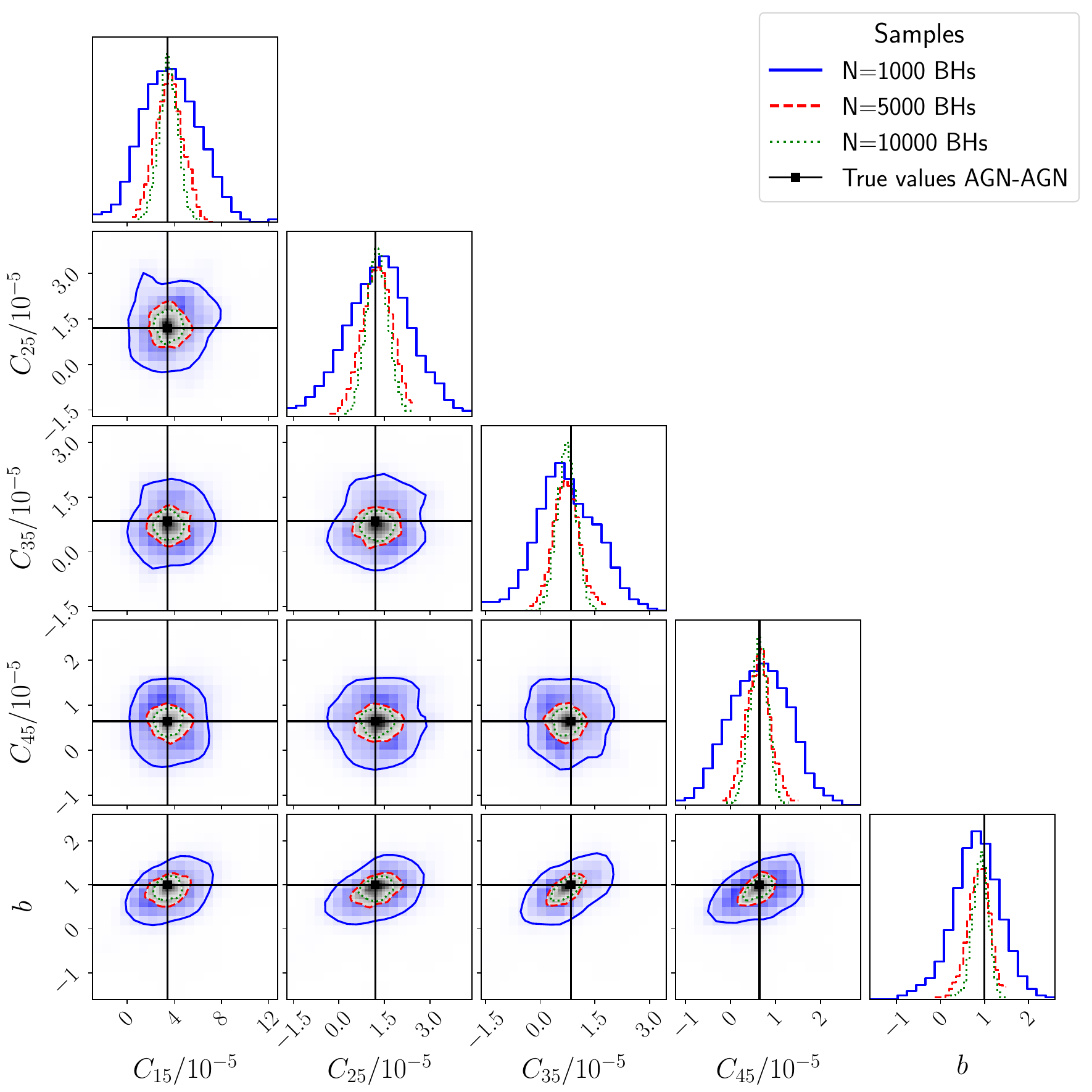}
    \caption{$C_{\ell}$ distribution of GWs from AGN cross-correlated with AGN catalogs. The expected value of $C_\ell^{AGN,AGN}$ (with bin averaging) is shown in black.}
    \label{fig:GWAGN_AGN_CC}
\end{figure}

We again consider three different sizes of GW populations, $N_{\rm{BBH}}=1000$, $N_{\rm{BBH}}=5000$, and $N_{\rm{BBH}}=10,000$. For each of these, we compute the ACC of the resulting combined GW skymap comprising all of the localization contours, with our AGN catalog (from which our true GW locations were initially sampled from). We repeat this $500$ times for each $N_{\rm{BBH}}$ new GW populations. We consider the resulting distribution of the $C_\ell^{GW,A}$ coefficients and the relative bias $b$, computed in each simulation from the best fit value of $b$ that minimizes the log loss function:

\begin{equation}
    \log \mathcal{L}(b)= \frac{1}{2} \sum_{\ell, \ell'} X_\ell(b) \cdot  (\Sigma^{-1})_{\ell \ell'}^{A,GW} \cdot X_{\ell'}(b),
\end{equation}
\noindent
where $X_{\ell}(b)=C_\ell^{A,GW}-b \cdot C_\ell^{A,A}$, and $\Sigma_{\ell \ell'}^{A,GW}$ is given by Equation \ref{eq:Covariance_matrix}. The expected relative bias is $b=1$, since the sampled true GW locations are drawn from the AGN catalog.

The result is shown in Figure \ref{fig:GWAGN_AGN_CC}. As expected, the distribution of $C_\ell^{GW,A}$ for each $N_{\rm{BBH}}$ is centered around the corresponding $C_\ell^{A,A}$ coefficient, similarly the distribution of the relative bias $b$ is peaked around the expected value of $b=1$. The distributions become narrower as $N_{\rm{BBH}}$ increases: the quality of clustering information improves with more detections. We observe that the bias introduced by removing events with low SNR is small enough as to not significantly alter the expected ACC. This demonstrates the robustness of using the ACC between GW localizations and galaxy catalogs to determine the origin of a population of BBH events.

\section{Case of multiple sources of hierarchical mergers}

One drawback of the approach described in section \ref{sec:ACC_point_estimate} is that it only works when either the AGN channel, or another non-AGN channel, dominates the hierarchical merger rate. Here we demonstrate an approach that can determine the fraction of hierarchically merging BBHs from the different sources in a given sample.

\subsection{Determining the fractional contribution of AGN vs. non-AGN to the hierarchical merger population}\label{sec:Methods_fraction}

Consider the ACC of a catalog X with a catalog $\rm{Y}+\rm{Z}$, produced by combining the skymap of the $n_{\rm{Y}}$ objects in catalog Y and $n_{\rm{Z}}$ objects in catalog Z. It follows from the definition of the density contrast given in Equation \ref{eq:delta_def} that
\begin{equation}
    \delta^{\rm{Y}+\rm{Z}}(\theta,\phi) = f_{\rm{Y}} \cdot \delta^{\rm{Y}}(\theta,\phi) + f_{\rm{Z}} \cdot \delta^{\rm{Z}}(\theta,\phi),
\end{equation}
\noindent
where $f_{\rm{Y}} = \frac{n_{\rm{Y}}}{n_{\rm{Y}}+n_{\rm{Z}}}$ and $f_Z=\frac{n_{\rm{Z}}}{n_{\rm{Y}}+n_{\rm{Z}}}$ are the fraction of the combined $\rm{Y}+\rm{Z}$ catalog composed of the $\rm{Y}$ catalog and the $\rm{Z}$ catalog respectively. It follows then from the definition of the ACC in Equation \ref{eq:Cl_cross_def}, due to the linearity of the integral in Equation \ref{eq:a_lm def}, that
\begin{equation}
             C_\ell^{X,Y+Z} = f_Y \cdot C_\ell^{X,Y} + f_Z \cdot C_\ell^{X,Z}.
\end{equation}
We now apply this property to a gravitational wave catalog $GW=AGW+NGW$, with a fraction $f_A$ coming from localizations drawn from an AGN catalog, and $f_N$ drawn from a non-AGN catalog. We expect the ACC of the GW skymap with an AGN and non-AGN skymap to be distributed as follows:

\begin{multline}\label{eq:AGW_Cl_mix}
    C_\ell^{A,GW} \sim f_A \cdot C_\ell^{A,AGW} + f_N \cdot C_\ell^{A,NGW}
     + (1-f_A-f_N) \cdot \epsilon^A,
\end{multline}
\begin{multline}\label{eq:SGW_Cl_mix}
    C_\ell^{N,GW} \sim f_N \cdot C_\ell^{N,NGW} + f_A \cdot C_\ell^{N,AGW}
     + (1-f_A-f_N) \cdot \epsilon^N.
\end{multline}

To determine  $C_\ell^{A,AGW}$, $C_\ell^{N,NGW}$, $C_\ell^{A,NGW}$, and $C_\ell^{N,AGW}$ we can perform similar simulations as described in Section \ref{sec:cross_cor_sim}. Alternatively, we find that they are well modeled as normally distributed random variables: 
\begin{align}
C_\ell^{A,AGW} &\in \mathcal{N}(C_\ell^{A,A}, \Sigma_{\ell \ell}^{A,AGW}),\\
C_\ell^{A,NGW} &\in \mathcal{N}(C_\ell^{A,N}, \Sigma_{\ell \ell}^{A,NGW}),\\ C_\ell^{N,NGW} &\in \mathcal{N}(C_\ell^{N,N}, \Sigma_{\ell \ell}^{N,NGW}),
\end{align}
where the covariance matrices are obtained from Equation \ref{eq:Covariance_matrix}. If the catalog is composed only of objects from catalog A or N, then $f_A+f_N=1$. To account for the possibility of catalog incompleteness, we allow $1-f_A-f_N > 0$, with the additional contribution $\epsilon^A$ ($\epsilon^N$) coming from the shot noise found by cross correlating catalog A (N) with a uniformly sampled sky map. This noise term is modeled as a normal distribution $\mathcal{N}(0,\Sigma_\ell^{A,rGW})$, with $\Sigma_\ell^{A,rGW}$ found from Equation \ref{eq:Covariance_matrix} evaluated under the conditions that 
\begin{equation}
C_\ell^{CAT,CAT}=C_\ell^{GW,GW}=C_\ell^{GW,CAT}=0,
\end{equation}
and 
\begin{align}
N_{CAT}&=N_{AGN},\quad N_{GW}=N_{BBH}.
\end{align}

Then, given population fractions $(f_A,f_N)$ and measured cross-correlations $C_\ell^{A,GW}$ and $C_\ell^{N,GW}$, we define the residual terms

\begin{equation*}
        \Delta_{\ell}^{A,GW} \equiv C_\ell^{A,GW}-f_A \cdot C_\ell^{A,A}-f_N \cdot C_\ell^{A,N},
\end{equation*}
\begin{equation*}
        \Delta_\ell^{N,GW} \equiv C_\ell^{N,GW}-f_N \cdot C_\ell^{N,N}-f_A \cdot C_\ell^{A,N}.
\end{equation*}

The expected values of these quantities, by Equations \ref{eq:AGW_Cl_mix} and \ref{eq:SGW_Cl_mix}, are $\langle \Delta_{\ell}^{A,GW} \rangle = \langle \Delta_{\ell}^{N,GW} \rangle = 0$. Further, assuming independence of $C_\ell^{A,AGW}$, $C_\ell^{N,NGW}$, $C_\ell^{A,NGW}$, and $C_\ell^{N,AGW}$, the covariances are 

\begin{equation}
    \rm{cov} \left(C_\ell^{A,GW} \right)=f_A^2 \Sigma_{\ell \ell'}^{A,AGW} + f_N^2 \Sigma_{\ell \ell'}^{A,NGW} + (1-f_A-f_N)^2 \Sigma_{\ell \ell'}^{A,rGW} \equiv \Sigma_{\ell \ell'}^{A,GW},
\end{equation}

\begin{equation}
    \rm{cov} \left(C_\ell^{N,GW} \right)=f_N^2  \Sigma_{\ell \ell'}^{N,NGW} + f_A^2 \Sigma_{\ell \ell'}^{N,AGW} + (1-f_A-f_N)^2  \Sigma_{\ell \ell'}^{N,rGW} \equiv \Sigma_{\ell \ell'}^{N,GW}.
\end{equation}

\noindent

Consider the likelihood function for the cross-correlations $C_\ell^{A,GW}$ and $C_\ell^{N,GW}$, given known fractions $f_A$ and $f_N$, and parameters $\vec{p}=(C_\ell^{A,A},C_\ell^{A,N},C_\ell^{N,N})$ determined by the galaxy catalogs. We construct the Guassian likelihoods for the ACC with GW catalogs:

\begin{equation}
    \mathcal{L}(C_\ell^{A,GW}|f_A,f_N;\vec{p}) \propto \exp \left[    -\frac{1}{2}  \sum_{\ell, \ell'}\Delta_{\ell}^{A,GW} \cdot (\Sigma^{-1})_{\ell \ell'}^{A,GW} \cdot \Delta_{\ell'}^{A,GW} \right],
\end{equation}

\begin{equation}
    \mathcal{L}(C_\ell^{N,GW}|f_A,f_N;\vec{p}) \propto \exp \left[    -\frac{1}{2}  \sum_{\ell, \ell'}\Delta_{\ell}^{N,GW} \cdot (\Sigma^{-1})_{\ell \ell'}^{N,GW} \cdot \Delta_{\ell'}^{N,GW} \right].
\end{equation}

\noindent
where by independence we have 
\begin{equation*}
    \mathcal{L}(C_\ell^{A,GW},C_\ell^{N,GW}|f_A,f_N;\vec{p}) = \mathcal{L}(C_\ell^{A,GW}|f_A,f_N;\vec{p}) \mathcal{L}(C_\ell^{N,GW}|f_A,f_N;\vec{p}).
\end{equation*}

\noindent
Given point estimates of $C_\ell^{A,GW}$ and $C_\ell^{N,GW}$, and a prior on the fractions $\pi(f_A,f_N)$, we can use Bayes theorem to obtain the joint posterior distribution of the fractions:

\begin{equation}\label{eq:posterior_mixture}
    p(f_A,f_N|C_\ell^{A,GW},C_\ell^{N,GW};\vec{p}) \propto \mathcal{L}(C_\ell^{A,GW},C_\ell^{N,GW}|f_A,f_N;\vec{p}) \pi(f_A,f_N)
\end{equation}

We can use the resulting maximum likelihood estimate -- $\hat{L}_{\rm{mix}}$ obtained from the fractional values $(\hat{f}_A,\hat{f}_N)$ maximizing the posterior in equation \ref{eq:posterior_mixture} -- to perform model comparison of different BBH origin channels. We employ the Akaike Information Crietrion (AIC) \citep{akaike1998information} to compare the hypothesis of the BBH merger population being sampled only from AGN, only non-AGN galaxies, a mix of AGN and non-AGN galaxies, or noise. 

\noindent
The AGN, non-AGN hypothesis, and noise hypothesis have the following likelihoods (obtained from Equation \ref{eq:posterior_mixture}): 

\begin{equation*}
\begin{split}
    \hat{L}_A &= p(f_A=1.0,f_N=0.0|C_\ell^{A,GW},C_\ell^{N,GW};\vec{p}), \\
    \hat{L}_{N} &= p(f_A=0.0,f_N=1.0|C_\ell^{A,GW},C_\ell^{N,GW};\vec{p}), \\
    \hat{L}_{n} &= p(f_A=0.0,f_N=0.0|C_\ell^{A,GW},C_\ell^{N,GW};\vec{p}).
\end{split}
\end{equation*}

The AIC for a model with $k$ free parameters and likelihood $\hat{L}$ is

\begin{equation}\label{eq:AIC}
    \rm{AIC} = 2 k - 2 \log{\hat{L}}.
\end{equation}

The mixed origin hypothesis has two free parameters $(f_A,f_N)$, and hence $k=2$ when computing the AIC from equation \ref{eq:AIC}.
Our other three hypotheses have no free parameters and hence $k=0$. The relative likelihoods of model $i \in \{\rm{mix},\rm{A},\rm{N},\rm{n}\}$ compared to the AIC of the favored hypothesis $\rm{AIC}_{\rm{min}}$ is found from:

\begin{equation}
L_\mathrm{relative} = \exp((\rm{AIC}_{min}-\rm{AIC}_\textit{i}) / 2),
\end{equation}

\subsection{Detecting a mixed catalog}\label{sec:appendix_fraction_measure}
Consider five simulated GW catalogs, each wtih $N_{BBH}=10,000$ BBHs with locations sampled from an underlying galaxy catalog with of AGN, $f_{AGN} \in \{0, 0.25, 0.5, 0.75, 1.0\}$. The remaining fraction of events $f_\mathrm{Non-AGN}=1-f_{\rm{AGN}}$ have their locations sampled from the non-AGN galaxy catalog. For each fraction $f_\mathrm{AGN}$, we determine the posterior distribution on the AGN and non-AGN fractions using \texttt{emcee} \citep{foreman2013emcee} . We use a Dirichlet prior with $K=3$ and $\alpha_1,\alpha_2,\alpha_3=1$, where the three parameters are $(f_A,f_N,1-f_A-f_N)$. For each simulation, we calculate the AIC to compare the hypothesis of pure AGN/non-AGN origin, a mixture model, and noise.

\begin{figure*}
    \centering
    \begin{subfigure}[b]{0.45\textwidth}
    \includegraphics[width=\textwidth,height=0.32\textheight,keepaspectratio]{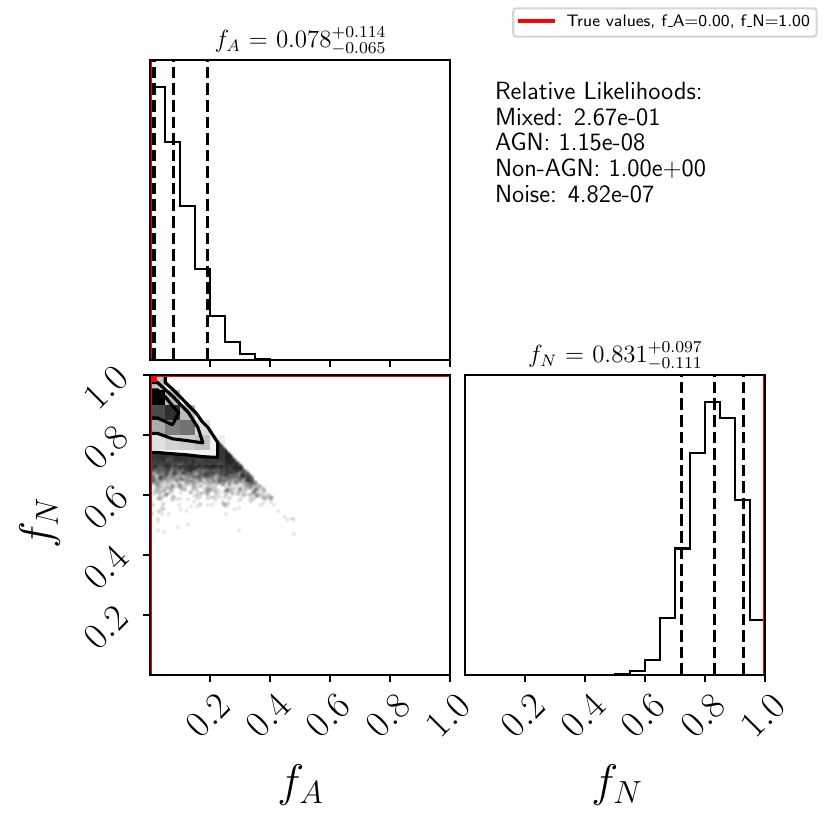}
        % \caption{cap}
    \end{subfigure}
    \hfill
    \begin{subfigure}[b]{0.45\textwidth}                \includegraphics[width=\textwidth,height=0.32\textheight,keepaspectratio]{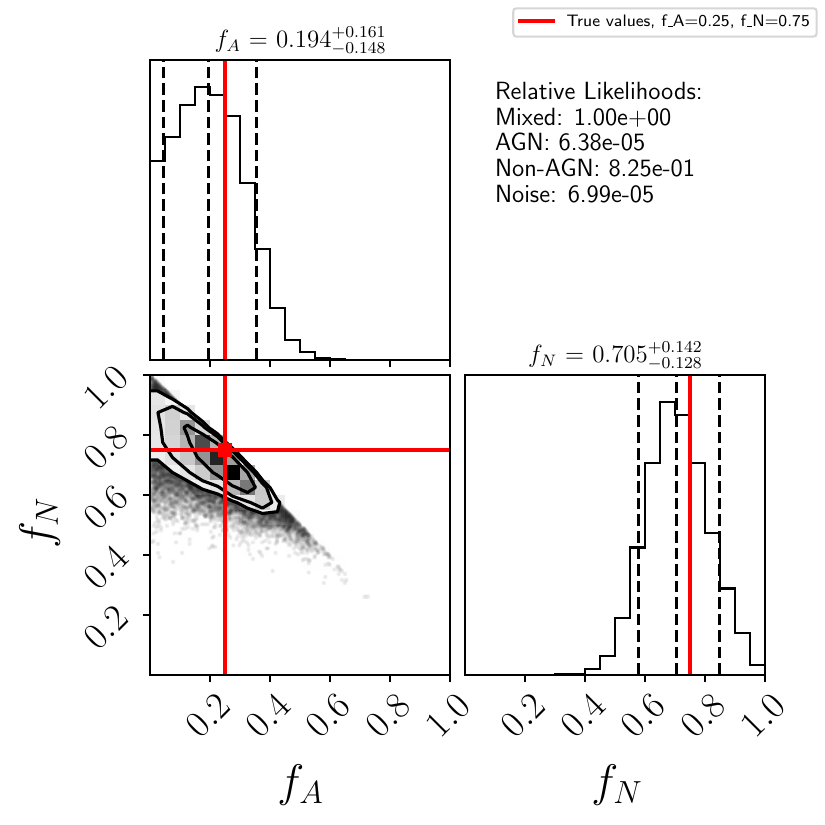}
        % \caption{.}
    \end{subfigure}
    \vfill
    \begin{subfigure}[b]{0.45\textwidth}
        \includegraphics[width=\textwidth,height=0.32\textheight,keepaspectratio]{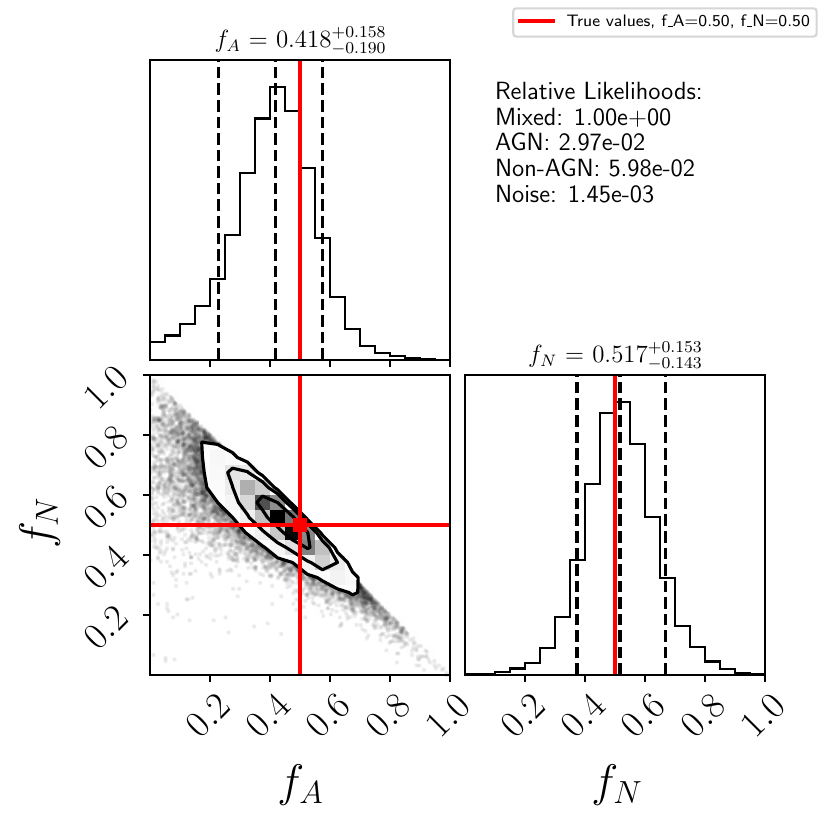}
        % \caption{}
    \end{subfigure}
    \hfill
    \begin{subfigure}[b]{0.45\textwidth}
        \includegraphics[width=\textwidth,height=0.32\textheight,keepaspectratio]{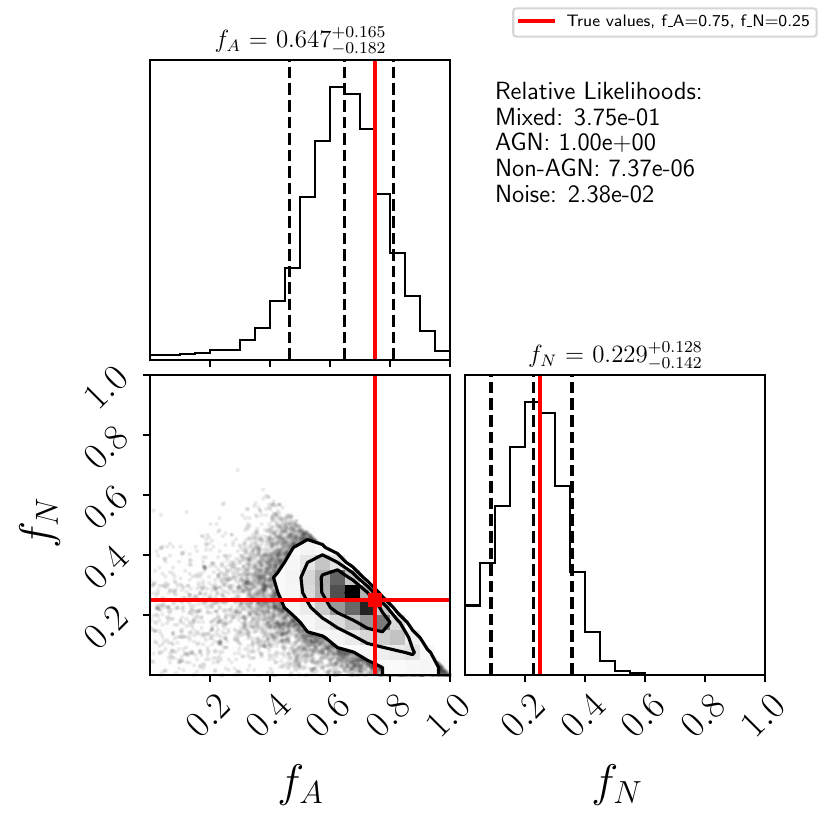}
        % \caption{}
    \end{subfigure}
    \vfill
    \begin{subfigure}[b]{0.45\textwidth}
        \includegraphics[width=\textwidth,height=0.32\textheight,keepaspectratio]{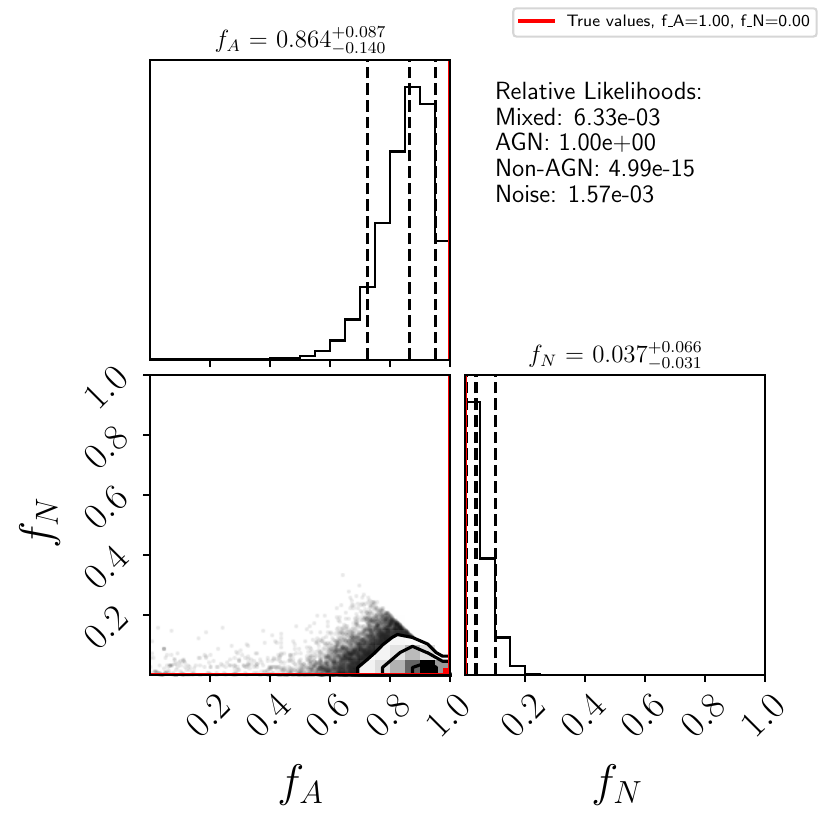}
        % \caption{}
    \end{subfigure}
    \caption{Estimates of the fraction of objects in a catalog that are sourced by AGN and non-AGN galaxies, for a single estimate with $N_{BBH}=10,000$, and varying fractions $f_{AGN} \in \{0.0, 0.25, 0.5, 0.75, 1.0$\} and $f_{N}=1-f_{AGN}$.}
    \label{fig:fraction_posteriors}
\end{figure*}

We show the result in figure \ref{fig:fraction_posteriors}, where each plot shows the posterior distribution of $(f_A,f_N)$ resulting for given input fractions (shown in red). Each plot also shows the relative likelihoods of the different origin scenarios obtained from the AIC, namely the pure AGN origin ($f_A=1$), the non-AGN origin ($f_N=1$), the mixed origin (where $f_A,f_N$ are the fractional values maximizing the likelihood), and the noise model (found by setting $f_A=f_N=0$). The AIC criterion does a good job at identifying the preferred origin of AGN or non-AGN - i.e. when $f_{N} = 1$ it predicts non-AGN origin, and when $f_{A} \in \{1,0.75\}$ it predicts AGN origin, and for all other fractions it predicts a mixed origin. Additionally, the posteriors on the fractions $(f_A,f_S)$ contain the input fraction within the $90\%$ confidence interval of the posterior in most cases, except for when $f_{A}=1$ or $f_{N}=1$. This serves as an example of the estimates on the AGN and non-AGN fraction of a single sample universe unlike our previous result, where we generate many realizations of possible GW catalogs).

\section{Discussion and conclusion}\label{sec:Discussion}
In this work, we have demonstrated that we can establish the origin of a population of hierarchical BBH mergers as being from an AGN- or non-AGN dominated channel through their angular cross-correlations with different putative host galaxy catalogs. Given the projected capabilities of the LIGO and Virgo interferometers at A\# sensitivity, we demonstrate that with $\mathcal{O}(5000)$ events, it is possible to distinguish which evolutionary channel dominates the hierarchical merger rate. These numbers and localization areas correspond to projected event numbers \citep{broekgaarden2024visualizing} and localization precision \citep{gupta2023characterizing}. Our method could be used to complement other sky localization techniques, models of parameter distributions for hierarchically merging BHs, and associations of GWs with AGN flares, in order to assess whether hierarchically merging BBHs form predominantly in AGN or non-AGN environments.

Some previous works studying the localization of BBH mergers to AGN focus on calculating the overdensity of AGN in the 90\% credible interval of the sky localization region for GW events \citep{bartos2017gravitational,veronesi2022detectability, veronesi2023most,veronesi2025constraining} as opposed to the statistical properties of the cross-correlations between catalogs. For example, \cite{veronesi2025constraining} rely on the relative sparsity of AGN compared to non-AGN galaxies to make assertions about host associations. In particular, only the highest luminosity, unobscured AGN with $L>10^{44.5} \rm{erg}\,\rm{s}^{-1}$ are considered. Catalogs of such objects have greater completeness at a given redshift when compared to catalogs that also include lower luminosity AGN. Consequently, the method of \cite{veronesi2025constraining} can only give information on the plausibility of association between GW events and the highest luminosity AGN. This method may be inadequate to explore the hypothesis that AGN are a dominant source of hierarchically-merging BBH: simulations suggest that higher luminosity AGN are not necessarily more likely hosts of these systems. For example, formation of migration traps is suppressed in high luminosity AGN \citep{grishin2024effect,gilbaum2024escape}. In support of the idea that lower luminosity AGN may be preferred environments in which this population of BBH form, a recent work by \cite{zhu2025evidence} claims that low luminosity AGN do contribute a significant fraction of mergers detected by LVK, on the order of $\sim 40\,\mathrm{per\,cent}$. In contrast to these two works, our approach makes no \textit{a priori} cut on the AGN luminosity to account for the apparent uncertainty in the relationship between AGN luminosity and ability to form BBH systems hierarchically. We also defer a thorough investigation on the impacts of catalog completeness on our method to future work. 

Our work follows the approach of considering statistical correlations in the spatial distributions of GWs and galaxies. Similar methods have been established previously -- including \cite{vijaykumar2023probing,raccanelli2016determining,oguri2016measuring,banagiri2020measuring,mukherjee2021accurate} -- with a focus on the application of using BBHs for cosmology. We adapt these methods to address the question of whether an already established to be hierarchically merging population of BBHs have formed in AGN disks or not. This is important for our understanding of the rates of different binary black hole formation channels, which are used to test models of black hole formation in interstellar environments, which themselves depend on various physical and astronomical effects that are uncertain.

This work was performed with the key assumption that the galaxy catalogs we use are complete, or close-to complete (in the sense that the angular cross-correlation can be accurately determined for the catalog). With forthcoming galaxy surveys such as the LSST \citep{LSST} and 4MOST \citep{4most}, obtaining galaxy catalogs that are close to complete for large portions of the sky will be an achievable expectation by the time GW detectors are operational at A\# sensitivity. Another limitation to applying our method to existing GW data arises from the poor sky localizations of the majority of events in GWTC-3, where the majority of events are localized to areas $\Omega\gg 10\rm{deg}^2$ \citep{abbott2023gwtc} meaning clustering information is lost, rendering this method insensitive to clustering on small size scales.

In principle, our method can be extended to include the complete three-dimensional GW localization posterior distributions, and precise galaxy redshifts. We ignore three-dimensional effects for two reasons: the paucity of AGN at low redshift means that these bins are shot-noise-dominated, and the potentially uncertain GW redshift posteriors make binning a significant challenge. Although inclusion of redshift information offers additional discriminatory power in principle, we find that our simple approach can also yield interesting results. We defer the inclusion of three-dimensional localizations and the challenge of determining appropriate binning in redshift space to future works.

We emphasise that our method is dependent on the observed sky localizations of a population of BBH events. In the limit of the Cram\'{e}r-Rao lower bound, applied through the FIF, the sky localization is dependent on the SNR of each individual event and therefore is implicitly dependent on the astrophysical parameters of the BBH mergers. For example, an excess of mergers with higher chirp masses, as expected for the AGN channel, could result in a population of events with higher SNRs and consequently smaller sky localization areas. We defer investigation of these biases to future works, although we anticipate that these biases would be small at a population level given the dependence of the SNR on other extrinsic parameters, e.g. the position of the event on the sky relative to the detector sensitivity, and the luminosity distance at which the events occur.

In conclusion, we demonstrate that cross-correlating the projected 2D localization regions for a putative catalog of hierarchically merging BBH with galaxy catalogs can shed light on the evolution of dynamically formed black hole binaries. This method could give key insights into the environments that give rise to gravitational wave events that occupy the upper mass gap - events that cannot be readily explained with most conventional stellar evolution models. 

\section*{Acknowledgements}
We are grateful to Simon Driver, Ilya Mandel, Bruce Gendre, Evgeni Grishin,  Eric Thrane, David Coward, Chris Blake and Christopher Lidman for helpful discussions. We thank the reviewer for their helpful comments. FHP is supported by a Forrest Research Foundation Fellowship. JWNM is supported by funding from the Australian Government Research Training Program. The authors are grateful for computational resources provided by the OzSTAR Australian national facility at Swinburne University of Technology. 

%%%%%%%%%%%%%%%%%%%%%%%%%%%%%%%%%%%%%%%%%%%%%%%%%%
\section*{Data Availability}

The code necessary to perform all the analysis carried out in this work is available upon reasonable request.

%%%%%%%%%%%%%%%%%%%% REFERENCES %%%%%%%%%%%%%%%%%%

\bibliographystyle{mnras}
\bibliography{example} % if your bibtex file is called example.bib

% Don't change these lines
\bsp	% typesetting comment
\label{lastpage}
\end{document}